# Opinion Divergence Reveals the Hierarchical and Overlapping Community Structure in Networks

Ren Ren, Jinliang Shao

*University of Electronic Science and Technology of China, Sichuan, China*

**Abstract**

Often exhibiting hierarchical and overlapping structures, communities or modular groups are fundamental and complex in network science. One of the most exploited tools to detect the mesoscopic structure is synchronization. Several phenomena including convergence rate and local convergence under constraints are studied to uncover the existence and features of communities. Here, employing a background of opinion dynamics, we study the divergence of agents' opinions, i.e. the state differences of nodes, in the progress of reaching global consensus and then reveal its connections with hierarchy and overlap in the modular structure. Furthermore, based on the obtained close relationships, a new method is proposed to identify hierarchical and overlapping communities, whose robustness and efficiency are validated via experiments on real and artificial networks. Both the connections and approaches provide a novel insight on the features of modular structures in networks.

*Keywords:* Complex Networks, Opinion Dynamics, Community Detection, Hierarchical Structure, Overlapping Structure

## 1. Introduction

Community structure, usually viewed as a group of nodes with strong internal but weak external connections [1], is a common feature found in actual observations of networks, for example, the karate network [2] and the citation networks [3]. Playing an important role on revealing underlying mechanisms of complex systems, community detection becomes a fundamental work in network analysis. Many metrics have been proposed to detect communities based on the inhomogeneity of edges' amount [1, 4, 5, 6], among which the modularity of Newman et al. is most studied [1]. Taking the metrics as evaluation indicators of communities' quality, researchers in several fields develop a number of static optimization approaches to solve the problem [7].

Apart from static optimization, detecting communities by running dynamical processes on the network is also an important methodology [8]. Therein, the random walk is widely used [9, 10, 11]. A typical example is hierarchical clustering based on node similarity measured by visiting frequency [9], whereas the partitions are finally determined by the fitness functions. Another network dynamics often exploited is synchronization [12, 13]. In [12], the authors simulate the synchronization of oscillators in the network, and then identify communities by inspecting the extent of local convergence of the nodes in different time periods. Later in 2011, Morarescu et al. employ an opinion dynamics system with time-decaying confidence fields to community detection [13]. The key point is that the confidence, which means the agents refuse to receive opinions far from theirs own greater than a bound, results in the existence of local converged groups, which are naturally identified as communities. Both the algorithms explore the relationships between local convergence in synchronization and community detection, offering insights different from optimizing benefits functions. Whereas, Fortunato points out that the synchronization method in Ref. [12] is instable [7]. Additionally, as shown in Morarescu's and other researchers' experiments, the algorithm with decaying confidence is also parameter sensitive [14]. Recently, Domenico studies the "diffusion distance" of dynamical processes such as random walks and synchronization, indicating that the new distance can reveal functional clusters in networks [15].

With the in-depth studies, researchers find that communities have their own complexity. One of the complex features is hierarchy [16], which means that communities are nested and can exhibit several levels in the network; Another is overlap [17], which implies that one node belongs to more than one community simultaneously. There are many works focusing on the hierarchical structure, most of which identifies communities at different meaningful levels by optimizing the ad hoc metrics [5, 4, 18]. Compared to hierarchy, overlap is even more appealing. Palla et al. firstly uncover the overlap of nodes in community structure by using the clique percolation method [17]. In [19], the authors provide a "functional definition" for overlapping nodes in the framework of synchronization, confirming the importance of overlap. Based on the assumption that edges are rarely overlapping, link clustering technique was developed in [5, 20]. In this kind of method, one link only belongs to one group, thus a node with several links can be in more than one group. Later, the highly overlapping phenomenon attracted widespread attention [4, 21], where the overlapping area (both nodes and edges) can be quite large. Local expansion approaches were developed to deal with this situation [4, 22], which relies on empirical fitness functions to decide the range of communities. Among all methods mentioned above, algorithms in Ref. [4, 5] are designed to detect hierarchical and overlapping communities meantime, unifying the two complexities in one framework.

Indeed, the emergence of communities in social networks is common, often owing to the disagreement of groups of agents, as the term "community" suggests. A detailed living example is the karate network with two communities reported by Zachary [2]. Because of the divergence on membership fees of the club, the members split into two groups and hence two communities emerge in the network. Democrats vs. Republicans is another good instance, we all know that Democrats and Republicans have divergence on lots of issues, and there are inevitable divergences inside each party, which may lead to small internal groups or in other words, hierarchical communities. Moreover, not all people entirely belong to one group, besides supporters and protesters, it is also common to see neutrals in social interactions. To understand the essence of these phenomena, many models are proposed to depict the evolution of opinions in networks [23, 24]. The fruitful studies inspire us to explore the complexities of communities focusing on the divergence in opinion dynamics, which gives a further comprehension of community structure.

In this paper, we propose a new method to detect hierarchical and overlapping communities at the same time in the framework of synchronization. We show that our approach is robust and effective via the experiments on artificial and real networks though no fitness functions are used. Simulating the opinion exchange progress on networks based on the DeGroot model [25], we analyze the divergence of agents during the progress, thus find a natural emergence of overlapping parts and hierarchical structure in communities, uncovering their distinguishable features in the framework. Compared to previous works mentioned above [15, 13], apart from revealing clusters in networks merely, we further explore the relationships between the differences of nodes' states and the complex structure of communities detailedly.

## 2. Preliminaries and Models

A system of some agents exchanging their opinions can be modeled by a network or graph $G = (V, E)$, where $V$ is the set of nodes or agents, and $E$ is the set of adjacent nodes' connections, namely edges or links. If some of the edges in $G$ are removed, which suggests that some agents refuse to communicate with others, the agents are divided into groups. Thus each group becomes a connected component, i.e. a subgraph where any two nodes are connected to each other by links. Let $A = (a_{ij})_{n \times n}$ denote the adjacency matrix of $G$ and $d_i = \sum_{j=1}^{n} a_{ji}$ is the degree of node $i$, where $n$ is the number of agents. Then the basic DeGroot model [25] in opinion dynamics can be expressed by

$$x(t+1) = D^{-1}Ax(t), \qquad (1)$$

where $t$ is the time step, $D = diag\{d_1, d_2, ..., d_n\}$ is the degree matrix, and $x(t) = (x_1(t), x_2(t), ..., x_n(t))^T$ is the vector of agents' opinions or states at time $t$. According to the equation above, each agent updates its opinion by the weighted average of the opinions of itself and its out-neighbors in the network. In this model, the agents reach a global asymptotic consensus as time steps increase, i.e. $\forall i, j, x_i(t) - x_j(t) \to 0, t \to \infty$. This model is widely accepted and has been extended to model other complex situations such as opinion dynamics with susceptibility to persuasion[23, 24].

In this article, we use the basic DeGroot model to simulate the procedure of agents' reaching consensus in the network. Because the convergence of the agents' states is asymptotic, it is meaningful to study the state differences in the progress, which can reflect the features of the network structure. More concretely, focusing on edges in the network, we study the differences or divergences of adjacent nodes. We define the opinion distance as follows:

**Definition 1.** for each pair of adjacent nodes $i$ and $j$, i.e. $a_{ij} > 0$, the opinion distance between $i$ and $j$ is:

$$\Delta(e_{ij}) = \sum_{t=t_0}^{\infty} |x_i(t) - x_j(t)|, \qquad (2)$$

where $e \in E$, and $t_0$ is the initial time to add up the opinion distance, specially, $\sum_{t=0}^{\infty} |x_i(t) - x_j(t)|$ is the total opinion distance of node $i$ and $j$. Usually we use $t_0 >= 5$ to avoid the impact of different initial values in the classical linear system Eq. (1). We define opinion distance for every pair of

adjacent nodes, which means each edge corresponds to a $\Delta$, so we give the definition as a function of an edge.

When $t \to \infty$, $x_i(t) - x_j(t) \to 0$, and in the community detection problem, it is unnecessary to use the accurate value of $\Delta$. In our algorithm, we use the estimated value $\delta$:

$$\delta(e_{ij}) = \sum_{t=t_0}^{t_1} |x_i(t) - x_j(t)|. \quad s.t. \ Var(x(t_1)) < \xi, \tag{3}$$

where $\xi$ is the threshold controlling the extent of the convergence, and $\xi = 10^{-12}$ in our experiments on community detection.

The opinion distance of two adjacent agents reflects their divergence in the process of reaching an agreement. The larger opinion distance, the harder to achieve consensus. Considering the situation in the real life, two persons with more divergence on something are more likely to be in the different groups. It is reasonable to deduce that: *An edge with high opinion distance have more possibility to be intra-community than inter-community.*

## 3. Algorithms

*3.1. Community Detection*

Based on the models and definitions in Sec. 2, we detect community structure by removing edges with high opinion distances. The detailed steps are stated in Algorithm 1. The key step in our algorithm is identifying which edges to remove at each iteration. Usually, internal edges are more than external edges, moreover, since that a community is a group of agents with similar opinions, the change intensity of internal opinion distances is not too large. Therefore,

$$\frac{\max\{\delta(e_{in})\} - \min\{\delta(e_{in})\}}{|\{e_{in}\}|} \ll \frac{\max\{\delta(e_{out})\} - \min\{\delta(e_{out})\}}{|\{e_{out}\}|}, \tag{4}$$

where $e_{in}$ is an internal link and $e_{out}$ is an external link of communities. The inequality suggests that, when sorted ascending, the external opinion distances increase much more quickly than those within groups as the order indexes increase. Now we give an alternative technique to identify the turning point of $\bar{\delta}$'s growth rate in the sorted sequence in Algorithm 2 and Fig. 2a. We emphasize that this technique is alternative in our algorithm, other methods to discover the turning point are also acceptable.

**Algorithm 1:** Community detection

**Input**  : The network $G$, the terminating threshold $\epsilon_1, \epsilon_2$.
**Output:** The communities.

1. Edge removal iteration step $N = 0$;
2. **repeat**
3.    **for** *each unfinished component C (G included)* **do**
4.       **if** *C is new emerged or $C = G$* **then**
5.          $N_C = 0$; Etimate $t_1$ according to Eq.(1) and Eq.(3);
6.       **end**
7.       Remove the hanging edges in $C$ recursively;
8.       **for** $\tau = 1$ *to* $T$ **do**
9.          Initialize $x(0) \in [0,1]^n$ randomly;
10.          Calculate $\delta_\tau$ and $x(t_1)$ according to Eq.(1) and Eq.(3);
11.          The variance of $x(t_1)$: $\sigma_\tau^2 = Var(x(t_1))$.
12.       **end**
13.       $\overline{\delta} = \sum_{\tau=1}^{T} \delta_\tau / T$, $N_C = N_C + 1$, $\overline{\sigma}_{N_C} = \sum_{\tau=1}^{T} \sigma_\tau / T$.;
14.       Sort the edges by their corresponding $\overline{\delta}$ ascending, denote the sorted sequence by $\{e_1, e_2, ..., e_m\}$, where $m$ is the number of undeleted edges;
15.       Find the turning point $\beta$ satisfying $\overline{\delta}(e_k)$ increases sharply as $k$ increases when $k \geq \beta$ using Algorithm 2. Mark the edges $\{e_k | k \geq \beta\}$;
16.       Edge removal: delete the marked edges in a descend order if the edge is not hanging;
17.       **if** $(m - \beta)/m < \epsilon_1$ *OR* $\overline{\sigma}_1/\overline{\sigma}_{N_C}, \overline{\sigma}_1/\overline{\sigma}_{N_C-1}, \overline{\sigma}_1/\overline{\sigma}_{N_C-2}$ *all less than* $\epsilon_2$ **then**
18.          Mark $C$ as "finished";
19.       **end**
20.    **end**
21.    $N = N + 1$;
22. **until** *all of the components are "finished"*;
23. Recover the removed links within the finished components and the pre-removed hanging edges.

**Algorithm 2:** Identifying the turning point

**Input** : The sorted sequence $\{(k, \overline{\delta}(e_k))\}_{k=1}^m$
**Output:** the turning point $\beta$ of $\overline{\delta}$'s growth rate.

1 Link $(1, \overline{\delta}(e_1))$ and $(m, \overline{\delta}(e_m))$ with a straight line $ax + by + c = 0$;
2 $\beta = \arg\min_k ak + b\overline{\delta}(e_k) + c$, namely, among the points below the line, find the point $(\beta, \overline{\delta}(e_\beta))$ whose distance to the line is the maximal;
3 Output $\beta$.

Another important issue is the stopping conditions of edge removal fore components. Apart from a trivial condition where the edges to be deleted is fewer than a threshold, we also provide another condition where the opinions at time step $t_1$ of the group are not sufficiently close with each other for several continuous edge removal iterations. We suggest the threshold $\epsilon_2 \leq 10^{-3}$. See Fig. 9c for an example of this condition.

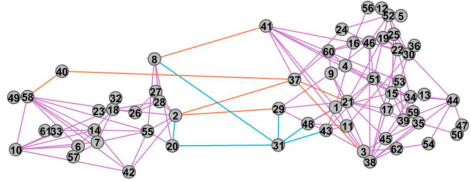

(a) 2 components of Dolphins

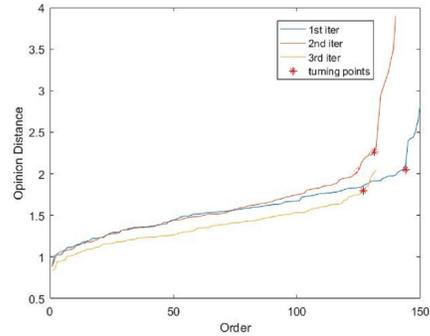

(b) opinion distance in iterations

Figure 1: Dolphins are divided into 2 connected components after 3 iterations, orange edges are marked in the first iteration, blue ones in the second iteration, and in the last iteration only 2 edges are marked, thus no more iterations.

In order to explain our algorithm more clearly, we take the famous Dolphin social network [26] as an example (Fig. 1). As it is shown, the orange edges are marked in the first iteration and blue edges are marked in the second iteration. After the edge removal operation in Step 5, the dolphins are divided into two groups corresponding to the ground truth. Note that the opinion distance between node 40 and 58 is less than that between node 40 and 37, so node 40 belongs to the left community in the figure, and for the same reason, node 31 belongs to the right community.

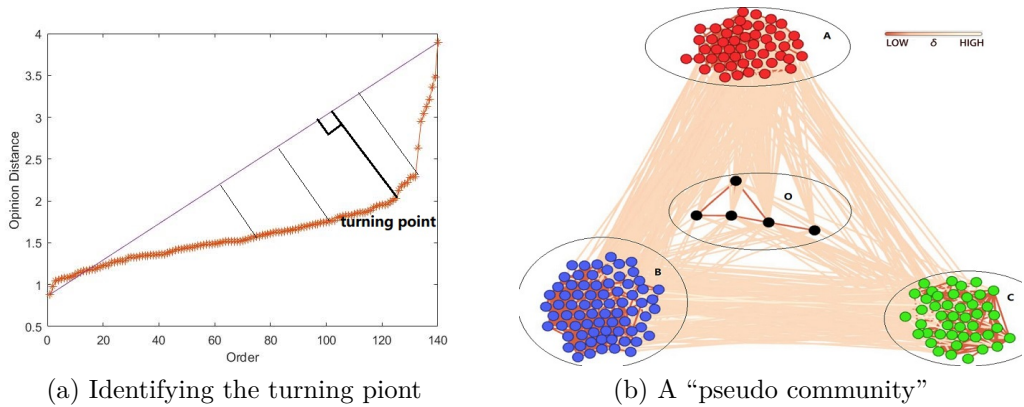

(a) Identifying the turning piont  (b) A "pseudo community"

Figure 2: (b): The deeper color of an edge, the smaller of its opinion distance. The nodes of Part O have only 5 internal links but many external links. O is a typical "pseudo community", all of the nodes in O belongs to community A, B and C at the same time.

*3.2. Connections with the spectra method*

In this subsection, we show the connections between our method and the spectra method [27] to provide more support for our community detection algorithm. Let us extend Eq. (2) first. As a row-stochastic matrix, $D^{-1}A$ has $n$ real eigenvalues, $D^{-1}A = U\Lambda U^{-1}$, where $\Lambda$ is the diagonal matrix of eigenvalues, $\Lambda = diag\{\lambda_1, \lambda_2, ...\lambda_n\}$ and $U = (u_{ij})_{n\times n}$ is the matrix whose columns are eigenvectors. Notice that $U^{-1} = U^T(D^{1/2})^T D^{1/2}$, thus $D^{-1}A = U\Lambda(U^T(D^{1/2})^T D^{1/2})$, then we have

$$\begin{aligned}
\Delta(e_{ij}) &= \sum_{t=t_0}^{\infty} |x_i(t) - x_j(t)| \\
&= \sum_{t=t_0}^{\infty} \left| \sum_l \sum_{k=2}^{n} \lambda_k^t u_{ik} u_{kl}^{inv} x_l(0) - \sum_l \sum_{k=2}^{n} \lambda_k^t u_{jk} u_{kl}^{inv} x_l(0) \right| \\
&= \sum_{k=2}^{n} \left( \left| \frac{\sqrt{d_k}\lambda_k^{t_0}}{1-\lambda_k} \right| |u_{ik} - u_{jk}| \left| \sum_l \sqrt{d_l} u_{lk} x_l(0) \right| \right)
\end{aligned} \qquad (5)$$

where $\lambda_k \in \Lambda$ is the *kth* largest eigenvalue of $D^{-1}A$, $\lambda_1 = 1$ and $u_{ij} \in U$, $u_{ij}^{inv} \in U^{-1}$. Note that in our simulation, the initial opinions $x(0) \in [0,1]^n$ of the agents are stochastic and independent with each other. Therefore, when the simulation is performed for enough times in the same network, the mean opinion distance between node $i$ and $j$ is,

$$\overline{\Delta(e_{ij})} = \sum_{k=2}^{n} \left( \left| \frac{\sqrt{d_k}\lambda_k^{t_0}}{1-\lambda_k} \right| |u_{ik} - u_{jk}| \left| \sum_l \sqrt{d_l} u_{lk} \overline{x_l(0)} \right| \right)$$

$$= \sum_{k=2}^{n} \left( \left| \frac{\sqrt{d_k}\lambda_k^{t_0}}{1-\lambda_k} \right| |u_{ik} - u_{jk}| \left| \sum_l \sqrt{d_l} u_{lk} \right| \overline{x_l(0)} \right)$$

$$= \sum_{k=2}^{n} \left( \left| \frac{\sqrt{d_k}\lambda_k^{t_0}}{1-\lambda_k} \right| |u_{ik} - u_{jk}| \left| \sum_l \sqrt{d_l} u_{lk} \right| \right) \overline{x_l(0)} \quad (6)$$

where $\overline{X}$ represents the expectation of the random variable $X$. Now we can define normalized opinion distance $\widetilde{\Delta}$ only decided by the graph $G$:

$$\widetilde{\Delta}(e_{ij}) = \frac{\overline{\Delta(e_{ij})}}{\overline{x_l(0)}} = \sum_{k=2}^{n} \left( \left| \frac{\sqrt{d_k}\lambda_k^{t_0}}{1-\lambda_k} \right| |u_{ik} - u_{jk}| \left| \sum_l \sqrt{d_l} u_{lk} \right| \right). \quad (7)$$

Considering the spectra method in [27], the authors embed the nodes in the network into a Euclidean space of $K-1$ dimension generated by $D^{-1}A$'s eigenvectors of top $K-1$ eigenvalues ($K$ is the number of communities), each node corresponds to a $K-1$ dimensional vector. After that they employ traditional clustering methods like K-means to cluster the embedded nodes and then get the corresponding communities. As shown in our equation 7, when $\widetilde{\Delta}(e_{ij})$ is large, $|u_{ik} - u_{jk}|$ is probably large for some $k$'s, which means node $i$ has a long distance with node $j$ in the Euclidean space, thus they are more likely to belong to different clusters. This connection offers an evidence for the basis of our algorithm that node $i$ and $j$ belong to different communities if $\Delta(e_{ij})$ is large enough.

## 4. Hierarchical and Overlapping Communities

### 4.1. Hierarchical Structure

It is natural to extend Algorithm 1 to hierarchical community detection because our edge removal operation from top to down has already form a dendrogram of subgraphs. We record the times of edge removal from initial graph $G$ to a component $C$ via $N$, and the life time of $C$ from its emergence to split via $N_C$ in the algorithm. We see the subgraphs with few life time (less than 3 for example) as unstable components, which is a interim status owing to the inaccuracy of identifying inter-group edges at current iteration. See an instance of the unstable component in Fig. 9a. With unstable ones excluded, the stable subgraphs emerge after similar number of iterations can be viewed as communities of the same level. In detail, if $C_1, C_2$ emerge when $N = n_1, n_2$, both $C_1, C_2$ are stable, and $|n_1 - n_2|/N_s$ is less then a threshold ($N_s$ is the value of $N$ when Algorithm 1 stops), then $C_1$ and $C_2$ are at the same level. The different emergence times reflect the heterogeneity of opinion distances, thus the level of hierarchical communities corresponds to the extent of opinion divergence among agents in opinion dynamics framework.

*4.2. Overlapping Structure*

Our algorithm detects communities based on the fact that internal opinion distances are less than external ones distinctly. In some networks, we find that even with all of the internal edges recovered, some components we get are loosely within-linked. Surely, those loose groups satisfy the mentioned basis of our method now that they are recognized. However, these groups should not be identified as independent communities because their internal edges are much fewer than external edges. Opposite to real communities, we call these loose groups "pseudo community" in this article. See the illustration of a "pseudo community" in Fig. 2b.

We simply use the proportion $f^C = K_{out}^C/K_{in}^C$ to judge a component is pseudo or real, where $K_{out}^C$ and $K_{in}^C$ are the total external and internal degrees of nodes in a component $C$ in the initial network. A group whose $f^C$ is higher than a threshold is not a community, and due to the large opinion distances with other real communities, it is not a part only belonging to one community either. Therefore, the group can only be seen as an overlap part of real communities, containing several nodes and edges. As for single overlapping vertices, they should have similar opinion distances with nodes of several communities. To decide the ascription of single overlapping nodes and pseudo communities, we design a score to describe the preferences of a single node to real communities, then the preferences of a pseudo community can be calculated by averaging the scores of its nodes. The preference of a single node to component $C$:

$$S_i^C = \frac{\sum_{j \in C \cap N_i} a_{ij}/\overline{\delta}^{(0)}(e_{ij})}{\sum_C \sum_{j \in C \cap N_i} a_{ij}/\overline{\delta}^{(0)}(e_{ij})}, \quad (8)$$

where $C$ is a real or pseudo community, $N_i$ is the set of neighbors of node $i$, $\overline{\delta}^{(0)}(e_{ij})$ is the opinion distance before edge removal iterations. The larger weight and smaller opinion distance of each link, the closer the connection is, hence we use the sum of $a_{ij}/\overline{\delta}^{(0)}(e_{ij})$ as the numerator and the denominator makes the score satisfy the normalization condition $\sum_C S_i^C = 1$. Given the score of preferences, we can assign the overlaps to real communities. See Algorithm 3 for details.

---

**Algorithm 3:** Overlapping community detection

**Input** : The components $\{C\}$ acquired by Algorithm 1
The Graph $G$, the threshold of a pseudo community $\gamma$
Opinion distance before any edge removal $\overline{\delta}^{(0)}$

**Output:** Overlapping memberships of each node $i$ to real communities $S_i^{RC}$

1 for each $C$, calculate $f^C = K_{out}^C / K_{in}^C$, if $f^C > \gamma$, mark $C$ "pseudo";
2 **for** *each node i in pseudo communities* **do**
3     for each real community $RC$, calculate $S_i^{RC}$
4 **end**
5 **for** *each pseudo community PC* **do**
6     for each component $C$, calculate $S_{PC}^C = \sum_{i \in PC} S_i^C / |PC|$
7 **end**
8 for each real community $RC$, $S_{RC}^C = 1$ if $C == RC$ else 0;
9 **for** *each node i in real communities* **do**
10     for each $C$, calculate $S_i^C$, if $C$ is pseudo, $S_i^{RC} = S_i^{RC} + S_i^C S_C^{RC}$ for each $RC$
11 **end**
12 **for** *each node i in G* **do**
13     Normalize $S_i^{RC}$ so that $\sum_{RC} S_i^{RC} = 1$, delete $S_i^{RC}$ if $S_i^{RC}$ is rather small (for example, less than 0.2), normalize $S_i^{RC}$ again
14 **end**

The score is only an indicator to decide the assignment of overlapping parts, while the judgment of overlaps is determined by the special pseudo communities, whose emergence suggests the features of highly overlapping community structure in our opinion dynamics framework. For both real and pseudo communities, the opinion distances of intra-group edges are less than those of inter-group edges. While the difference is that pseudo communities are loosely within-linked and densely external-linked, opposite to real ones, thus identified as overlapping parts. Here, "loose" and "dense" are relative, which implies the ambiguity of communities and overlapping parts. Indeed, in the real world, neutrals themselves can reach consensus more easily with each other. Moreover, if the neutrals connected densely enough, a new community may emerge. Typical instances in this situation include the establishment of a new party and the emergence of a new interdisciplinary.

## 5. Results

*5.1. Real-world Networks*

To demonstrate the effectiveness of our methods on real-world networks, we test our algorithm on Karate [2], Dolphins [26], Football (corrected) [10, 28] and Lesmis [5]. See our results of these networks in Fig. 1 and Fig. 3-5. We get completely correct results on Dolphins, almost correct partitions with only one misplacement on Karate, while other algorithms tend to identify more communities in these two networks [29, 30, 31, 11]. As for Football, we detect the communities consisting of non-independent nodes correctly, and find there is a small community of five independent nodes. Besides binary networks, our algorithm is also suitable for weighted networks, our partition is almost the same with that of classical Louvain method [29]. By the contrast of networks before and after edge removal, it is obvious that we delete the inter-community links without destroying the groups in community detection.

*5.2. Synthetic Networks*

LFR benchmarks [32, 33] are widely used to evaluate the performance of community detection algorithms, whose key parameter determining the complexity of community detection is the mixing parameter $\mu$, meaning the fraction of a node's edges shared with members of other groups. Using generally accepted Normalized Mutual Information (NMI [7]) as the performance measure, we generated four LFR networks with $\mu = 0.1, 0.3, 0.5, 0.6$ as benchmarks (Table 1). The comparison between ours and several other famous methods is shown in Table 2, which indicates that our method is robust and efficient. Moreover, it is noteworthy that our methods is excellent to reveal the true relationships in real-world social networks.

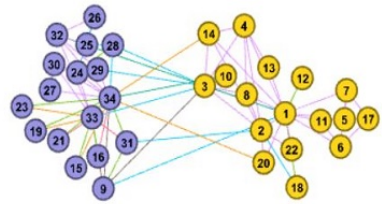
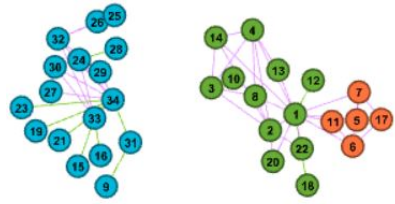

(a) Original Karate network  (b) Communities in Karate

Figure 3: The obtained communities in Karate, the nodes' colors in (a) show the ground truth, and the colors in (b) show the partition of the Louvain algorithm.

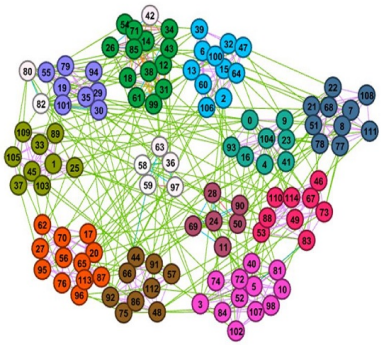
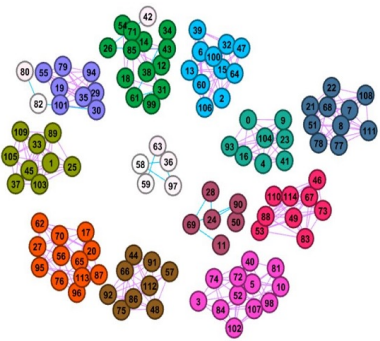

(a) Corrected Football network  (b) Identified communities in Football

Figure 4: Results on the corrected Football network, and nodes of the same color are in the same real-world communities except that the white nodes are independents.

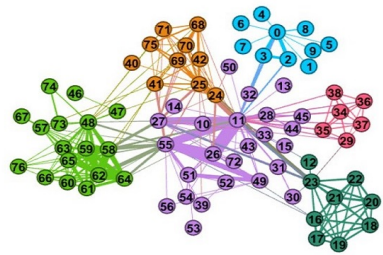
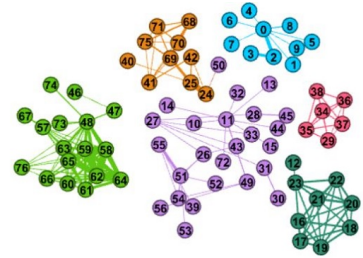

(a) Original weighted Lesmis network  (b) Identified comunities of Lesmis

Figure 5: The nodes of the same color are in the same community detected by Louvain.

| **Networks** | $\mu$ | $N$ | $k$ | $\tau_1$ | $\tau_2$ | $C_{min}$ | $C_{max}$ | $N_c$ |
|---|---|---|---|---|---|---|---|---|
| LFR1 | 0.1 | 1000 | 20 | 2.5 | 1.5 | 20 | 200 | 13 |
| LFR2 | 0.3 | 1000 | 20 | 2.5 | 1.5 | 20 | 200 | 20 |
| LFR3 | 0.5 | 1000 | 30 | 2 | 1.2 | 40 | 200 | 11 |
| LFR4 | 0.6 | 1000 | 30 | 2 | 1.2 | 40 | 200 | 13 |

Table 1: The parameters of 4 LFR benchmark graphs. $\mu$: the mixing parameter, $N$: number of nodes, $k$: average degree of nodes, $\tau_1$: negative exponent of degree's power:law distribution, $\tau_2$: negative exponent of the community size's power-law distribution, $C_{min}$: the minimum size of communities, $C_{max}$: the maximum size of communities, $N_c$: number of communities.

|  | Our Method | | | Infomap [30] | | Louvain [29] | | WalkTrap [9] | |
|---|---|---|---|---|---|---|---|---|---|
|  | C | Iter | NMI | C | NMI | C | NMI | C | NMI |
| LFR1 | 13 | 1 | **1.00** | 13 | 1.00 | 13 | 1.00 | 13 | 1.00 |
| LFR2 | 20 | 2 | **1.00** | 20 | 1.00 | 20 | 1.00 | 20 | 1.00 |
| LFR3 | 11 | 3 | **1.00** | 11 | 1.00 | 11 | 1.00 | 11 | 1.00 |
| LFR4 | 13 | 7 | 0.88 | 425 | / | 13 | **0.98** | 13 | 0.95 |
| Karate | 2 | 2 | **0.84** | 3 | 0.70 | 4 | 0.59 | 2 | 0.73 |
| Dolphins | 3 | 2 | **1.00** | 4 | 0.50 | 4 | 0.48 | 2 | 0.82 |

Table 2: Comparison of several methods on some networks. C: number of detected communities; Iter: number of iterations; NMI: normalized mutual information.

*5.3. Highly Overlapping Networks*

The Facebook network of Caltech [34] is a typical highly overlapping social network, which forms several groups due to the different background of students. We detect 7 communities in the network, whose extent of overlapping is displayed via the nodes' colors in Fig. 6a, where the deeper a node's color (both warm and cold) is, the more communities the node belongs to. The vertices belong to 5 communities at most and 1.62 in average, thus the network is highly overlapping. Note the community in warm colors emphasized at the right-upper corner of the figure, it shows that not only the frontier, but most of its agents are overlaps. This pattern gives a instance of many overlaps forming a remarkable community, which is not hard to understand in real world, for example, with the development of the inter-disciplinary, a new independent discipline may appear. In fact, Yang and Leskovec reported the presence of dense overlaps and the indistinguishability from community cores [21]. For comparison, the GCE method [35] reveals 9 communities with default parameters, where the nodes belong to 1.34 communities in average. We ignore the two smallest communities of size 17 and 38, viewing all their nodes as overlaps.

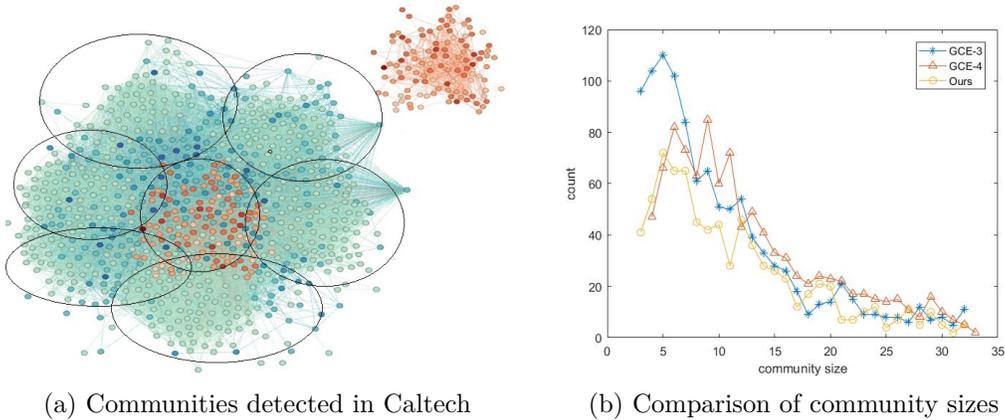

(a) Communities detected in Caltech  (b) Comparison of community sizes

Figure 6: (a): 7 overlapping communities were identified in the Facebook100-Caltech network. The deeper color of a node, the more communities it belongs to. Therein, the community in warm colors are extremely highly overlapping. (b): This figure shows that 3-clique GCE recognizes too many communities with few members and 4-clique GCE cannot detect triangle communities while our algorithm is suitable for this situation.

Apart from dense networks, our algorithm has superior on sparse networks with small communities. In the citation network reported by Newman in 2001 [37], we find 41 communities with 3 members and 54 communities with 4 members. However, for classical clique expansion methods, to get triangular groups, users must take 3-cliques as seeds, causing the loss of accuracy. Indeed, it is recommended to use 4-cliques as seeds in [35], ignoring 3-member groups. See the number of part of the communities detected by different methods and parameters (from size 3 to size 35) in Fig. 6b.

Last, we use LFR benchmark graph for overlapping community detection to test our algorithm and make comparison with other state-of-art methods.

The benchmark graph is with parameters $\mu = 0.4$, $o_n/n = 0.1$ (the proportion of overlapping nodes), and $o_m = 3$ (the maximum number of groups one node can belong to). We view the components whose $f^C > 1.5$ as overlapping parts, and get a partition whose NMI with the ground truth is 0.90. As shown in Table 3, our method is the best among the classical approaches. Therefore, our algorithm for overlapping community detection is effective and explicable.

| Methods | Ours | NDOCD | LC | OCG | CPM | GCE |
|---------|------|-------|------|------|------|------|
| NMI | **0.90** | 0.81 | 0.39 | 0.44 | 0.68 | 0.67 |

Table 3: Parameters of the benchmark: $n = 500$, $k = 25$, $k_{max} = 50$, $\mu = 0.4$, $\tau_1 = 2$, $\tau_2 = 1$, $c_{max} = 50$, $c_{min} = 10$, $o_m = 3$, $o_n = 50$ and $N_c = 22$.

*5.4. Hierarchical Networks*

To validate our algorithm on hierarchical community detection, we apply it to a real-world social network Jazz [38]. The splits of graphs emerge at the 2nd, 8th and 14th iteration, and no other subgraphs exists. Since the interval of iterations between divisions are negligible, it is reasonable to see each split as one level of hierarchy. See the structure of Jazz in Fig. 7.

As for the synthetic hierarchical networks, we introduce a benchmark proposed by Fortunato et al [18]. For authority, we generate a benchmark graph with suggested parameters whose structure is displayed in Fig. 8. After 5 iterations, the standard deviation of $x(t_1)$ increased from $10^{-7}$ to $10^{-2}$. The smallest community at the high level was separated. The remaining graph was divided into two groups just at the next iteration. Therefore, the three communities we got so far are of the same level. Continuing dealing with the three subgraphs independently, we eventually get all the communities at the lower level, shown in Fig. 9. Fig. 9c displays the final state of one of the small communities, in which situation the standard deviation of $x(t_1)$ is large ($10^{-3}$), satisfying the terminating conditions, hence there are only two layers of nested communities.

We find out all the communities of all levels with few misplaced nodes, revealing the hierarchy of the network in our framework. For a better understanding of hierarchical structure, we draw the histograms grouping edges according to their opinion distance at some iterations in Fig. 10, now that the communities have already been identified. As shown clearly, the links between communities of upper level have larger median opinion distance than links between communities of lower level, while the internal edges' opinion distances have the smallest mid-value. The difference shown in the histograms, on the one hand, is an evidence of the validity of our method, confirming that edges between upper-level communities are prior to be removed, on the other hand, it provides a new insight of hierarchy. Communities at different levels have distinct extent of internal and external divergence. More detailedly, the opinion divergence between upper communities are larger than that between lower groups, and a large divergence between lower groups may be a small divergence relatively in an upper group, which is in accordance with our experience on government organizations, knowledge systems, etc.

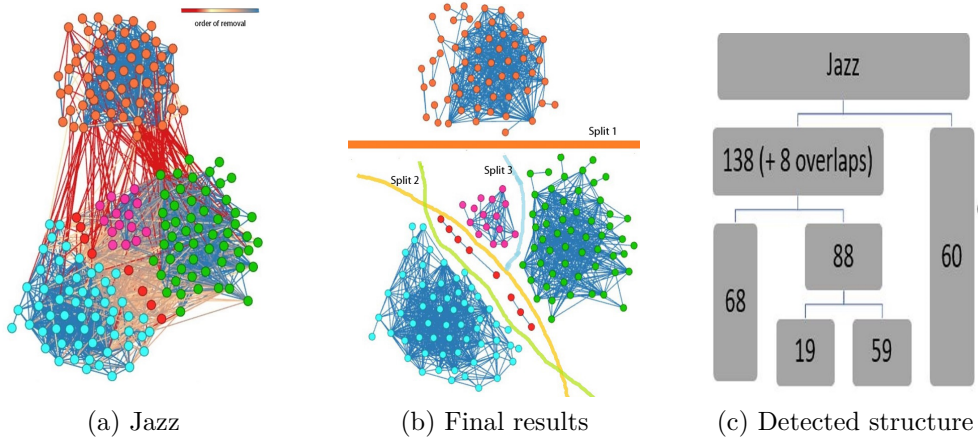

(a) Jazz  (b) Final results  (c) Detected structure

Figure 7: (a) The ranking colors of links denote the order of removal. The redder, the earlier; the bluer, the latter. (b) The nodes in deep red are two pseudo communities. (c) The detected hierarchical structure of Jazz, the numbers in the dendrogram are community sizes.

## 6. Conclusion

In this paper, we study the relationships between community structure and the differences of agents' states of the classical DeGroot model in opinion dynamics. Owing to the asymptotic convergence of agents' opinions, we define an "opinion distance" for each edge as the total differences of its two endpoints. We uncover that the opinion distance of an inter-community edge is usually higher than that of an intra-community edge. Moreover, when inspecting networks with hierarchical communities, the edges between nested communities of different layers have different levels of opinion distance. Therefore, either nested or not, a community is a group with small internal and large external opinion distances in our framework.

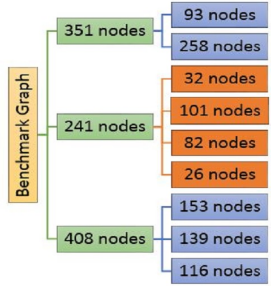 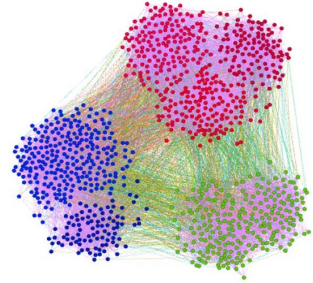

(a) The structure of the benchmark graph    (b) Orginal benchmark graph

Figure 8: The details of the benchmark graph with 1000 nodes and 7167 edges. The degrees of nodes are between minimum 5 and maximum 70, subjecting to power-law distribution with exponent 2.

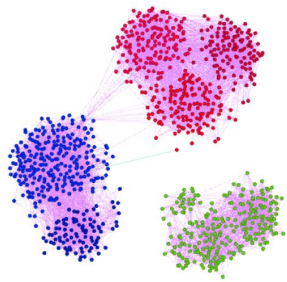 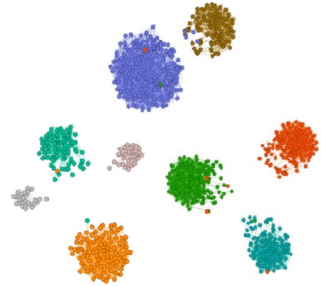 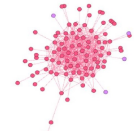

(a) First group is sperated    (b) Final results    (c) Terminating state

Figure 9: (a) The communities of the upper level. (b) The communities of the lower level. (c) Applying the algorithm on one of the small communities, the nodes appear radially. There is no sub-groups any more.

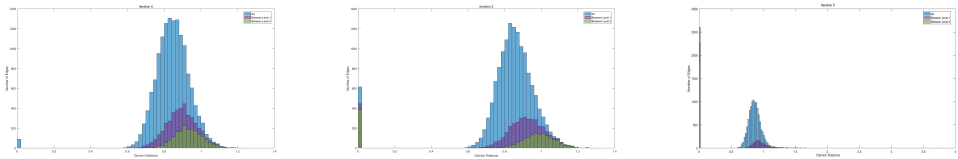

Figure 10: Some histograms of edges and opinion distance at two levels after each iteration. The edges with zero opinion distance are removed or hanging. Only hanging edges are removed at Iteration 0.

However, we also find some groups whose external opinion distances are higher than internal ones, while their internal edges are much fewer than external ones, which is opposite to real communities. Our analysis shows that they should be identified as overlapping parts of relevant communities. This discovery suggests the feature of the overlapping structure and the fuzziness between communities and dense overlapping parts.

The findings allow us to develop a practical community detection algorithm, which unifies hierarchy and overlap in one framework on the basis of opinion dynamics. Experiments on artificial and real-world networks show that our approaches are efficient and robust. Therefore, the obtained discoveries and detection methods offer new insight on community structure and its close relationships with social opinion evolution phenomena.